\documentclass[english,aps,prl,twocolumn,showpacs,superscriptaddress,footinbib]{revtex4-1}
\usepackage[T1]{fontenc}
\usepackage[latin9]{inputenc}
\usepackage[a4paper]{geometry}
\usepackage{babel}
\usepackage{amsmath}
\usepackage{amssymb}
\usepackage{graphicx}
\usepackage[unicode=true,pdfusetitle,
 bookmarks=true,bookmarksnumbered=false,bookmarksopen=false,
 breaklinks=false,pdfborder={0 0 1},backref=false,colorlinks=false]
 {hyperref}

\makeatletter

\@ifundefined{textcolor}{}
{%
 \definecolor{BLACK}{gray}{0}
 \definecolor{WHITE}{gray}{1}
 \definecolor{RED}{rgb}{1,0,0}
 \definecolor{GREEN}{rgb}{0,1,0}
 \definecolor{BLUE}{rgb}{0,0,1}
 \definecolor{CYAN}{cmyk}{1,0,0,0}
 \definecolor{MAGENTA}{cmyk}{0,1,0,0}
 \definecolor{YELLOW}{cmyk}{0,0,1,0}
}

\usepackage{diagbox}
\usepackage{makecell}
\usepackage{upgreek}
\def\bibsection{%
   \section*{References}%
   \@nobreaktrue}

\makeatother

\begin{document}

\title{Reply to comment on \textquotedblleft Spatial Coherence and Optical
Beam Shifts\textquotedblright{}}

\author{W. Löffler}

\email{loeffler@physics.leidenuniv.nl}

\affiliation{Huygens--Kamerlingh Onnes Laboratory, Leiden University, P.O. Box
9504, 2300 RA Leiden, The Netherlands}

\author{Andrea Aiello}

\affiliation{Max Planck Institute for the Science of Light, Günther-Scharowsky-Straße
1/Bldg. 24, 91058 Erlangen, Germany}

\affiliation{Institute for Optics, Information and Photonics, Universität Erlangen-Nürnberg,
Staudtstr. 7/B2, 91058 Erlangen, Germany}

\author{J. P. Woerdman}

\affiliation{Huygens--Kamerlingh Onnes Laboratory, Leiden University, P.O. Box
9504, 2300 RA Leiden, The Netherlands}

\maketitle

In \cite{wang2013comment}, Wang, Zhu and Zubairy repeat their previous
claim \cite{wang2008} that the spatial Goos-Hänchen (GH) shift happening
at total internal reflection at a dielectric-air interface depends
on the spatial coherence of the incident beam. This contradicts our
theoretical and experimental findings \cite{loffler2012sc}. Here,
we show that the apparent disagreement between their numerical simulations
and our results occurs only in a parameter range where the concept
of a spatial beam shift is invalid, and that therefore their claim
is inapplicable. We clarify this by discussing two key issues. 

First, Wang \emph{et al.} observe their effect only if the beam half-opening
angle $\theta_{0}$ is large compared to the difference between the
incident angle $\theta_{inc}$ and the critical angle $\theta_{crit}$:
$|\theta_{inc}-\theta_{crit}|\lesssim\theta_{0}$. In this case, part
of the beam is actually only \emph{partially} reflected, which is
in obvious contradiction to the statement in \cite{wang2008} that
they aim to investigate the spatial beam shift under \emph{total }internal
reflection. The Gaussian Schell Model (GSM) beam half-opening angle
$\theta_{S}$ is given by \cite{mandel1995} 
\[
\theta_{S}^{2}=\frac{2}{k^{2}}\left[\left(\frac{1}{2\sigma_{S}}\right)^{2}+\left(\frac{1}{\sigma_{g}}\right)^{2}\right].
\]

For instance, let us consider a case addressed by Wang \emph{et al.}
\cite{wang2013comment} (Fig.~2 therein, envelope waist $\sigma_{S}=100\,\mu$m,
transverse coherence length $\sigma_{g}=6.8\,\mu$m), which we will
refer to as ``beam \emph{A}'' below. In this case, the half-opening
angle is $\theta_{S}=1.3^{\circ}$. If the beam is incident close
to the critical angle, about half of the beam is partially reflected
(see inset Fig.~1). By definition, this is not a spatial beam shift
anymore; a spatial beam shift requires that, upon reflection, the
plane-wave components pick up only a phase varying linearly with the
angle (see Ref.~\cite{gragg1988} for an accessible discussion thereof).
This becomes obvious by realizing that in the case of Wang \emph{et
al.}, the reflected beam is no longer propagation invariant, in fact,
part of the beam experiences an \emph{angular} GH shift \cite{chan1985}.
This can be seen in Fig.~1, which shows the cross section of the
reflected beam \emph{A} during propagation. We also stress that the
beam deformation happening in such a case demands careful treatment
of the \textquotedblleft beam position\textquotedblright , such as
via the centroid or $1^{\mathrm{st}}$ order moment of the intensity
distribution; the determination of beam position numerically via the
peak position as done by Wang \emph{et al.} \cite{wang2008,wang2013comment}
is arbitrary. 

\begin{figure}[b]
\includegraphics[width=1\columnwidth]{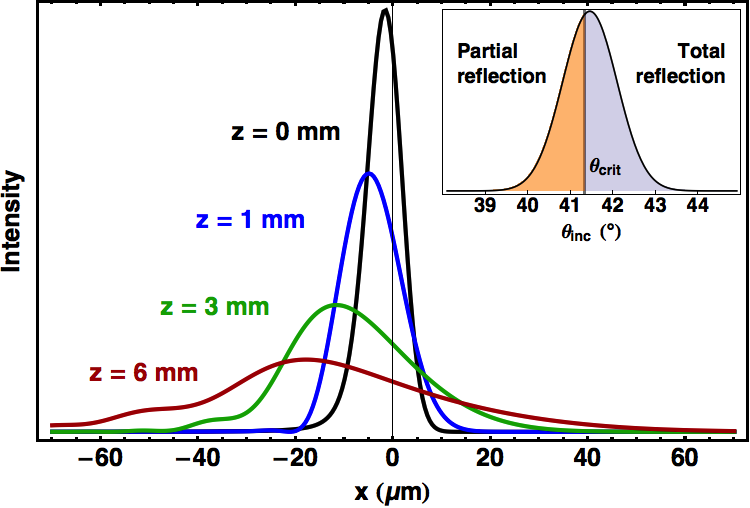}\protect\caption{Reflection of the \emph{p}-polarized beam \emph{A} at around the the
critical angle ($\theta_{inc}=41.45^{\circ},\,\theta_{crit}=41.34^{\circ}$,
$\theta_{S}=1.3^{\circ}$) at different propagation distances; the
evident beam deformations are due to the mixed spatial--angular character
of the shift. Inset: Fourier spectrum of the incident beam.}
\end{figure}

As a second issue, discussion of beam shifts only makes sense if the
incident and reflected field are actually ``beams'' according to
the paraxial wave equation. This has been worked out by Mandel and
Wolf for the case of GSM beams (\cite{mandel1995}, p.~278 Eq.~5.6-73).
As an example, some of the GSM beam parameters used by Wang \emph{et
al.} in \cite{wang2008} fulfill the beam condition (e.g., for $\sigma_{gy}/\sigma_{y}=0.1$
at $\sigma_{y}=50\,\mu$m; $\sigma_{y}$ and $\sigma_{gy}$ correspond
to the usual GSM parameters $\sigma_{S}$ and $\sigma_{g}$ transformed
into the interface plane). However, some other clearly violate the
beam condition, e.g., for $\sigma_{gy}/\sigma_{y}=0.01$. Actually,
the field in the latter case would be highly divergent with a half-opening
angle $\theta_{S}\approx45^{\circ}$. In this regime (and even more
for the case of a ``point like source'' \cite{wang2013comment}),
reflection at a planar interface requires full vector treatment, which
is not provided by the scalar numerical simulations of Wang \emph{et
al.} \cite{wang2008,wang2013comment}. Basically, in strongly divergent
fields, the polarization cross-spectral density function $\overleftrightarrow{\mathbf{W}}$
does not factorize into a polarization and a spatial part anymore,
see \cite{wolf2003} for a discussion. This is because such strongly
focussed beams cannot be homogeneously polarized since transversality
of the plane-wave components must be maintained; this is called spin-orbit
coupling of light.

We conclude that, as reported in \cite{loffler2012sc}, the spatial
GH shift of a \emph{bona fide} beam is \emph{not} affected by spatial
coherence of the incident beam.

\pagestyle{empty}\bibliographystyle{apsrev4-1}
\bibliography{bibliography}

\end{document}